\documentclass[usenatbib,conference]{basi}
\usepackage[T1]{fontenc}
\usepackage[british]{babel}
\usepackage[varg]{txfonts}
\usepackage{rotating}
\usepackage{dcolumn}
\begin{document}
\title[Spectral stellar libraries and the VO]{Spectral stellar libraries and the Virtual Observatory}
\author[E.~Solano]%
       {E.~Solano$^{1,2}$\thanks{email: \texttt{esm@cab.inta-csic.es}} \\
       $^1$Centro de Astrobiolog\'{\i}a (INTA-CSIC), Departamento de Astrof\'{\i}sica,
       P.O. Box 78, 28691 Villanueva de la Ca\~{n}ada, Madrid, Spain \\
       $^2$ Spanish Virtual Observatory
       }

\pubyear{2014}
\volume{10}
\pagerange{\pageref{firstpage}--\pageref{lastpage}}

\date{Received --- ; accepted ---}

\maketitle
\label{firstpage}

\begin{abstract}
In this presentation I will briefly describe the main characteristics of the Virtual Observatory as a 
research infrastructure for astronomy, identifying those fields in which VO can be of help for the 
community of spectral stellar libraries. 
\end{abstract}


\section{Introduction}\label{s:intro}

A typical working scenario in modern astrophysics is that in which astronomers make queries to 
databases with contents of very different nature (scientific publications, observational data, physical 
parameters,...), and the result of these queries is downloaded through high-capacity networks for their 
subsequent analysis with high-performance computers.

This framework, familiar to all of us, is, nevertheless, far from an ideal working 
scenario as it is affected by a 
number of issues that limit the efficient exploitation of the contents of astronomical archives and 
services. The major drawbacks are the following:

\begin{itemize}
 \item Data discovery: The first problem is associated to the data discovery process, that is, the process 
 of finding where the data I am interested in are. A significant number of spectral stellar 
 libraries are easily accessible at different internet sites or available on the excellent web maintained by 
 D. Montes\footnote{www.ucm.es/info/Astrof/invest/actividad/spectra.html}. Nevertheless, questions on the 
 completeness of the compilation, the efficient search of the available resources and their operability 
 remain open.
 
 \item Data access: The second weak point is related to the heterogeneity in the ways of accessing the astronomical 
 services, mainly due to the lack of a standard access protocol. For instance, there are services which provide 
 stellar spectra through a web page which allows queries by different parameters 
 (e.g. MILES\footnote{http://miles.iac.es}). Other resources provide a web page with a list of links 
 to download the files (e.g. ASTRAL\footnote{http://casa.colorado.edu/~ayres/ASTRAL/}) or just a link from
 where a compressed tar file can be downloaded 
 (e.g. PHOENIX\footnote{http://phoenix.astro.physik.uni-goettingen.de/}). In this scenario, if the user has
 to query a relatively large number of services, the amount of time invested in learning 
 different user interfaces, access and download procedures makes the data access a very 
 inefficient process.
 
 \item Data representation: Not only the access to the spectral stellar libraries but also the format of 
 these data collections is heterogeneous. Sometimes, the 
 datasets are delivered in FITS format (e.g. ASTRAL), sometimes as zip files containing datasets in ASCII
 format with a header (e.g. SpeX\footnote{http://www.BrownDwarfs.org}) or, like the Kurucz 
 models\footnote{http://kurucz.harvard.edu/grids.html}, with 
 a header that is the FORTRAN programme required to read the data.  
 
 \end{itemize}
 
 The conclusion to all this is that, putting together pieces of information coming from different 
 astronomical resources is tedious, inefficient and very time-consuming, and clearly hinders the analysis 
 and exploitation of astronomical data.
 
 \section{The Virtual Observatory}\label{s:vo}
 
 Give a solution to the problems described in the previous section is the challenge facing the Virtual 
 Observatory (VO), an international 
 initiative that was born in 2000 with the objective to guarantee an easy and efficient access and 
 analysis of the information hosted in astronomical archives. There are twenty initiatives 
 distributed worldwide and grouped around IVOA\footnote{http://www.ivoa.net}, the International Virtual 
 Observatory Alliance.

 The solution proposed by the VO to solve the data discovery problem was the creation of
 a registry system. Registries \citep{2011arXiv1110.0513B} are sort of yellow pages, a mechanism with which VO applications can 
 discover and select the resources of interest (spectra, images or catalogues). 
 
 Regarding the problem of 
 data access, VO proposes a standard protocol to query spectroscopic data: the Simple Spectral Access 
 Protocol or SSAP \citep{2012arXiv1203.5725T}. According to it, all VO services providing spectra should be able to understand a query 
 formed by the URL of the service, the central position and the radius of the search cone. Theoretical 
 spectra constitute a particular case as the query by coordinates is meaningless. In this case, the 
 protocol establishes a server-application dialogue to identify the parameters contained in the models and 
 the range of allowed values.
 
 Finally, with respect to the problem of data description, VO proposes a standard data model. Data Models 
 are abstract representations that define a common framework for describing astronomical data. The VO 
 spectrum data model \citep{2012arXiv1204.3055M} contains 
 all the parameters that are necessary to fully describe spectroscopic data and defines the 
 dependences among them. Fields to include fundamental information to perform a scientific analysis 
 (statistical and systematic errors, data quality, aperture size,..) are contemplated in the data model.
 Credits are also taken into account. This is an important issue because, if they are
 not properly handled, this may discourage 
 data providers to expose their contents in the VO. The  
  spectrum data model includes different fields to properly propagate information both on the publisher 
  and the creator. Moreover, tools like VOSA\footnote{http://svo2.cab.inta-csic.es/theory/vosa/index.php} 
  provides by default a BibTeX file acknowledging properly 
  all the resources the researcher has made use of (catalogs, models,...) and that can be appended in a 
  straightforward way in the publication.
  
  The Virtual Observatory is not just standards. Standardization is necessary but it represents only the 
  first layer, the other two being analysis tools and VO science. It is important to stress that the 
  Virtual Observatory is a science-driven initiative whose ultimate goal is, with a few standards and a
  bunch of tools, to produce lot of VO-science, that is, science that cannot be done (or could be done 
  in a much less efficient way) without the Virtual Observatory.
 
\begin{figure}
\centerline{\includegraphics[width=11cm]{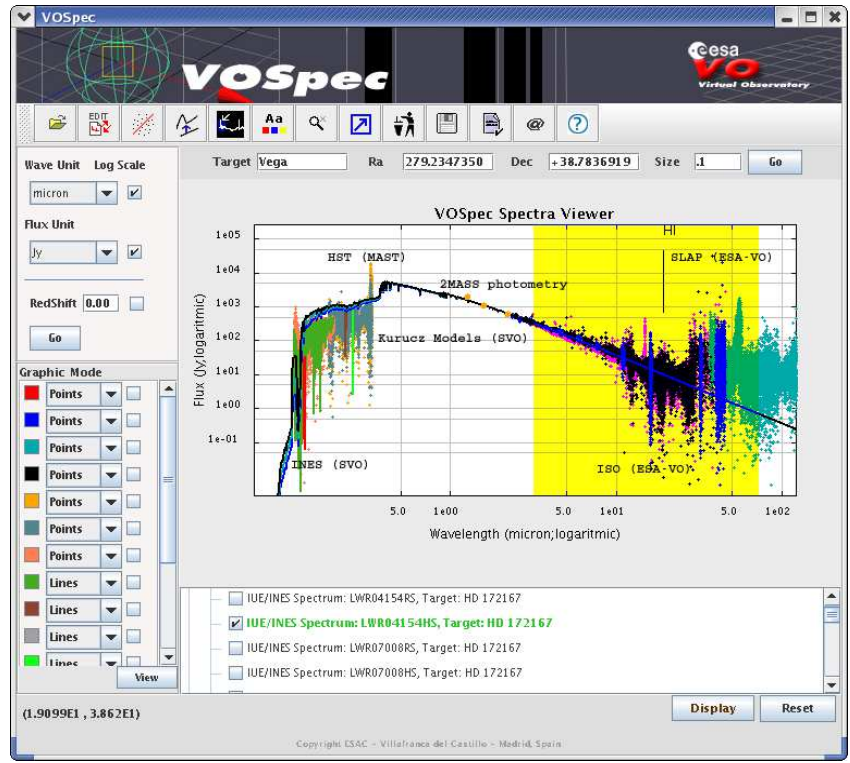}}
\caption{Generation of spectral energy distributions in the Virtual Observatory.}
\label{vospec}
\end{figure}

 Fig.~\ref{vospec} illustrates the power of the VO standardization. If we want to build the spectral energy 
 distribution of a given object using a VO tool like, for instance, 
 VOSpec\footnote{http://www.sciops.esa.int/index.php?project=SAT\&page=vospec}, the only thing we need to do 
 is to give the name of the object or its coordinates. The application will query the registry to look 
 for spectroscopic services. With the list of services provided by the registry, VOSpec will make the same 
 standard SSAP query to each of them. The services will return data in the same format and in compliance 
 with the spectrum data model. This will allow VOSpec to make basic operations like unit conversion and 
 show all the spectral pieces in the same scale. All this process is carried out in a transparent way 
 for the user.

\section{Publication of spectral stellar libraries in the VO}

There are different reasons why a spectral stellar library provider should publish their 
collections in the Virtual Observatory. Here I
will highlight some of them.

\begin{figure}
\centerline{\includegraphics[width=11cm]{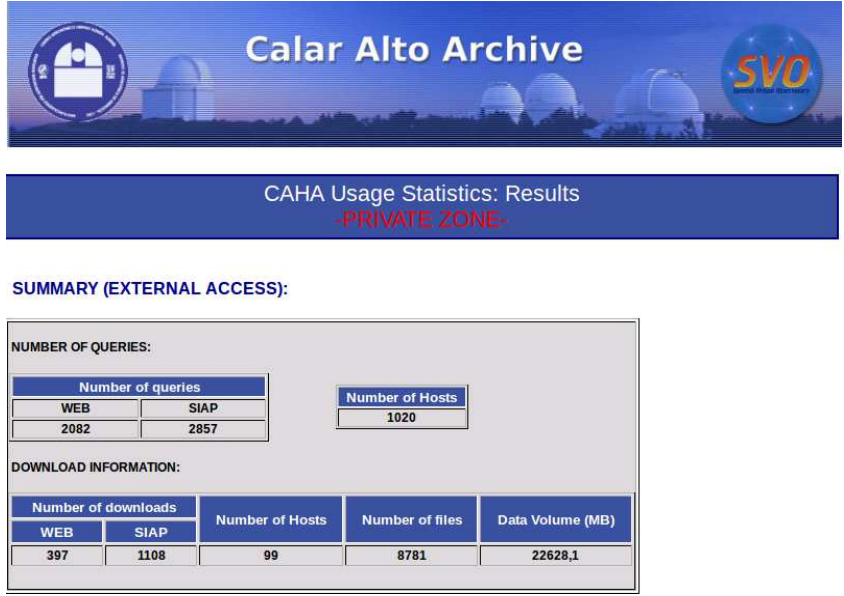}}
\caption{Usage statistics for the Calar Alto archive. We can see how the number of queries and downloaded 
files is larger using the VO access protocol than the web access.}
\label{calto}
\end{figure}

\begin{itemize}
\item Importance: The Virtual Observatory represents the framework where to carry out 
archive-related activities in Astronomy and as such has been recognized by 
Astronet\footnote{www.astronet-eu.org} in Europe and by the Decadal Survey in the 
US. 
\item Operational e-infrastructure: The Virtual Observatory is not just a nice idea or a prototype. Its 
operational phase started 
some years ago and many data centres (almost a hundred in Europe according to the last census) have already 
adopted the VO standards. 

\item Visibility: A good indicator of the success of a project is the amount of people not directly involved 
in it who have used the data. The Virtual Observatory is an excellent mechanism to enhance the visibility 
of an archive. Fig.~\ref{calto} shows the statistics on 
the SIAP (the protocol to access images in the VO) for the Calar Alto archive. We can see how, 
in 2012, 
both the number of queries and the number of retrievals has been larger using SIAP than using the standard 
web interface.

Spectral stellar libraries published in the VO will be easier accessed by the community which will 
eventually translate into a higher usage and more citations.

\end{itemize}

A major limitation in the publication of spectral stellar libraries in the VO is the fact that most
of the data providers are small teams, mainly focused on science, and that do not have 
either the time or the expertise or the willingness to learn how to design, develop, implement and maintain 
a VO-compliant archive.

To try to alleviate this situation, the Virtual Observatory has carried out a number of initiatives: A 
guide on how to publish data in the VO is available at 
the IVOA portal\footnote{http://wiki.ivoa.net/twiki/bin/view/IVOA/PublishingInTheVONew} and a number of 
workshops about this 
topic were held in the last years. In Europe, the Euro-VO\footnote{http://www.euro-vo.org} project provides
support on how to build a VO-compliant archive to those groups with the know-how necessary 
to maintain the service. Some national VO initiatives (e.g. SVO\footnote{http://svo.cab.inta-csic.es} or
GAVO\footnote{http://www.g-vo.org/}) also offer to host (with the appropriate credits) the data of those 
groups interested in having their collections in the VO but that do not want to be responsible of the 
archive.

\section{Usage examples}

\subsection{Comparison among synthetic collections}

We are now in the era of large sky surveys. They are powerful resources for research and discovery 
because of their data volume, homogeneity and depth. Spectroscopic surveys are, in particular, important 
because spectra are imprinted with information on the most fundamental stellar parameters.

Large surveys observe many objects of very different nature. The determination of 
the physical parameters for these objects requires the comparison with theoretical models or observational 
templates covering different regions of the parameter space. In particular, synthetic libraries have 
several advantages over observed spectra: they provide absolute fluxes and can be computed for 
any combination of stellar parameters.

Typically, large surveys (e.g. Gaia) make use of synthetic libraries that are not disjoint sets but they 
overlap
in different regions of the space of parameters. As the ingredients that form the libraries can
be 
different (different numerical approaches, different model atmospheres, different atomic and 
molecular 
line parameters,...), it is important to 
perform sanity checks to assess their coherence and identify potential systematic errors 
before deriving results from observational data. An standardized framework like the Virtual Observatory is 
the optimum scenario to carry out these sanity checks in an efficient way. 

\subsection{Enlargement of the covered spectral range}

 We can also take advantage of the Virtual Observatory to expand the spectral coverage of a 
 given stellar 
 library with spectra available in archives covering other wavelength ranges or to compare with 
 data
 in the same range for calibration purposes.
 
 Fig.~\ref{vospec} is an excellent example on how efficient the data discovery, access and retrieval is 
 for VO-compliant resources. 
 
 \subsection{Generation of synthetic photometry}

\begin{figure}
\centerline{\includegraphics[width=11cm]{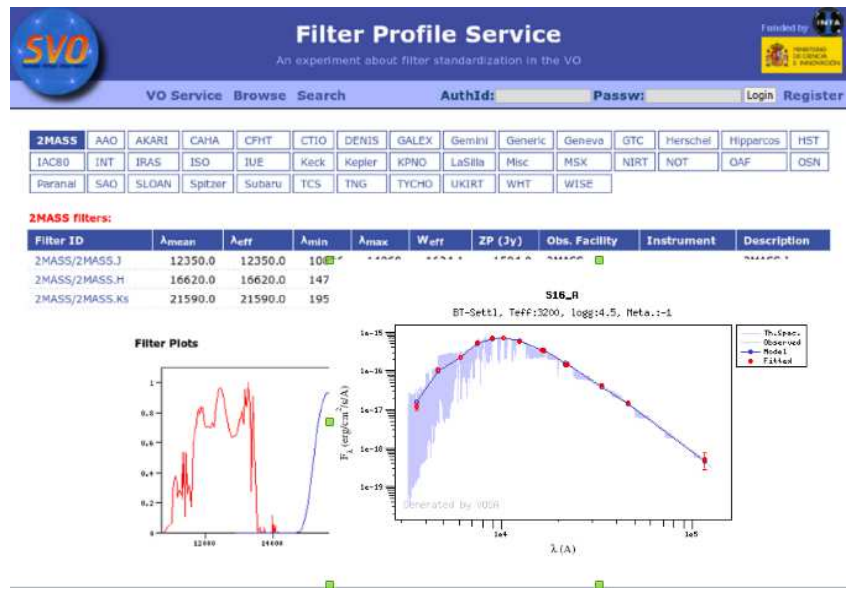}}
\caption{Outputs from the VO SED Analyzer (VOSA) and the Filter Profile Service. Both services have been 
developed by the Spanish Virtual Observatory.}
\label{vosa}
\end{figure}

Another typical task is the generation of synthetic photometry to compare with that obtained in
large area photometric surveys. VOSA \citep{2008A&A...492..277B} is a VO-tool to gather photometry from different VO services and 
compare them with the synthetic photometry obtained from different grids of theoretical models and 
observational libraries convolved with the photometric filters available in the SVO 
filter profile service\footnote{http://svo2.cab.inta-csic.es/theory/fps/index.php} to obtain 
the associated physical parameters (Fig.~\ref{vosa}). VOSA is a 
well-tested, robust tool with more than 200 users analyzing more than 165.000 objects (only in 2012) and 
around 35 refereed papers making use of VOSA since 2008.

\section{Future work}

The Virtual Observatory is an evolving infrastructure. Although it is producing science for some years,
there are still some lines of work that need to be improved. Among them, I will highlight three types of services:

\begin{itemize}
 \item Cutouts. 
 
 Asteroseismic studies based on spectroscopic data 
 typically demand the analysis of the variations with time of the line profiles in high 
 resolution spectra. In
 this case, the download of the whole echelle spectra for the subsequent cutout of the lines of 
 interest is not the best approach. Nevertheless, only very few VO services providing spectra implement this 
 capability.
 
 \item Spectral line identification
 
 Line lists are fundamental components for the computation of synthetic spectra and line 
 identification is a necessary step for all spectroscopic studies. In the Virtual Observatory there 
 exist an access protocol (SLAP, \citet{2011arXiv1110.0500S}) and a data model \citep{2011arXiv1110.0505O} for spectral 
 lines. Moreover, some tools like VOSpec implement functionalities to access different services 
 (e.g. NIST) and provide information on spectral lines. But the answer from these services includes all 
 the possible transitions in the wavelength range of interest, even though the influence of many of these 
 transitions in the observed spectrum is 
 completely negliglible . What a user would like to have is a service
 that, taking into account the physical parameters of the objects, 
 (\rm {T$_{eff}$}, \rm{logg}, \rm{[M/H]}), select only those lines with equivalent 
 widths (and/or central depths) above a given threshold. 
 
 VALD\footnote{http://www.astro.uu.se/~vald/php/vald.php} is a service that provides such a 
 functionality but VALD is not VO-compliant. It is now part of another European project called 
 VAMDC\footnote{http://www.vamdc.eu/}. Certainly, some liaison between VO and VAMDC in this field would be 
 very convenient. 
 
 \item Massive analysis
 
 The vast amount of data generated by the present and future spectroscopic surveys cannot be 
 analyzed 
 ``by hand'' but data mining techniques must be used instead. Gaia-ESO, SDSS-SEGUE, RAVE,..., 
 are examples of projects where automated tools for spectra classification and stellar 
 parameters determination will be necessary.
 
\end{itemize}

\section*{Acknowledgements}

This work has been done in the framework of the Spanish Virtual 
Observatory\footnote{http://svo.cab.inta-csic.es} 
supported from the Spanish MINECO through grant AyA2011-24052 and the CoSADIE FP7 project (Call 
INFRA-2012-3.3 Research Infrastructures, project 312559).

\end{document}